\documentclass[journal,onecolumn,10pt]{WSPLC_class}

\usepackage[left=2cm,top=1.78cm,right=2cm]{geometry}
\usepackage{graphicx}
\usepackage{verbatim}
\usepackage{amsmath}
\usepackage{subfig}
\usepackage{cleveref}
\begin{document}
\title{New Findings about Multi Conductor Noise in Narrow Band PLC}
\author{\IEEEauthorblockN{Davide Righini and Andrea M. Tonello}\\
\IEEEauthorblockA{Ecosys Lab, University of Klagenfurt, Austria\\ 
Email: {\{davide.righini, andrea.tonello\}@aau.at}}}
\maketitle

\section{Introduction}
The market projection for Internet of Things (IoT) systems is appealing for the Power Line Communication (PLC) technology. A massive growth and a pervasive presence of these IoT objects is expected in our everyday life. About 20 billion of IoT devices are forecasted to be installed in the next two years \cite{1,2}. PLC is a compelling technology that could become essential for certain IoT applications \cite{3}. For example, sensor nodes and actuators used in the Smart Buildings domain can be connected through PLC. In this case, the PLC network will work as a backbone for the in-building communication. PLC systems, in this domain, will help to solve or reduce the overcrowded-spectrum problem, exploiting the existing cable infrastructure.

Narrow band (NB) PLC is probably the most suitable type of PLC technology for IoT applications. The main reasons are: simple hardware, good coverage, appropriate bit-rate and low cost \cite{4}. Nevertheless, to guarantee reliable connectivity further analysis and improvements should be done. There are such as signal propagation discontinuities \cite{5}, high channel attenuation and noise \cite{6}.  
Despite the fact that PLC noise is a well-known problem, it still poses several challenges for this technology. This paper aims at adding some new consideration on this topic. Among the papers about PLC noise, \cite{7} presents a characterization on PLC noise types and \cite{8} describes an experimental characterization of the PLC noise at the source. More analysis on multi-conductor (MC) noise for broad-band PLC is presented in \cite{9,10,11}.

This paper describes new results on MC NB noise. The MC noise is measured as described in \cite{12} between the Live-Protective Earth (L-PE) and between Neutral-Protective Earth (N-PE) pairs (L-PE-N configuration), to exploit the symmetries of the channel. The analysis in this paper considers the whole PLC NB frequency spectrum that ranges from 3 kHz to 500 kHz. It considers the noise observed at the two output ports of a given outlet in an in-building network with three conductors.

\section{Noise characterization}
The PLC noise is characterized through an experimental measurement campaign. The noise signals are acquired in time domain with the usage of a digital storage oscilloscope (DSO) connected to a certain power grid outlet via a MIMO 2x2 NB coupler. The DSO is synchronized to the mains through a trigger device. The measurement setup parameters are fully described in \cite{12}.
\begin{figure}[!h]
	\centering
	\subfloat[]{{\includegraphics[width=8.5cm,height=5cm]{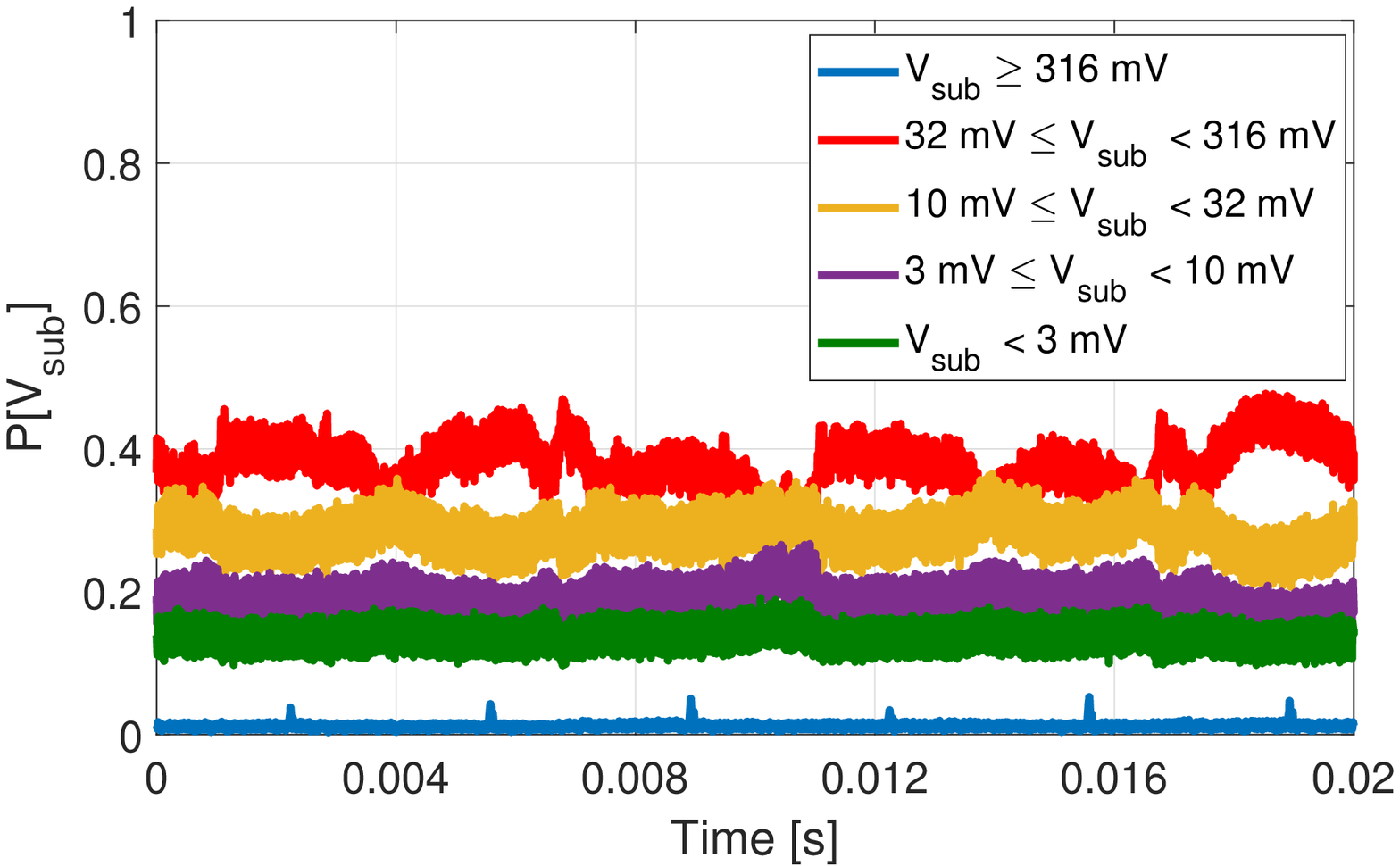} }}%
	\subfloat[]{{\includegraphics[width=8.5cm,height=5cm]{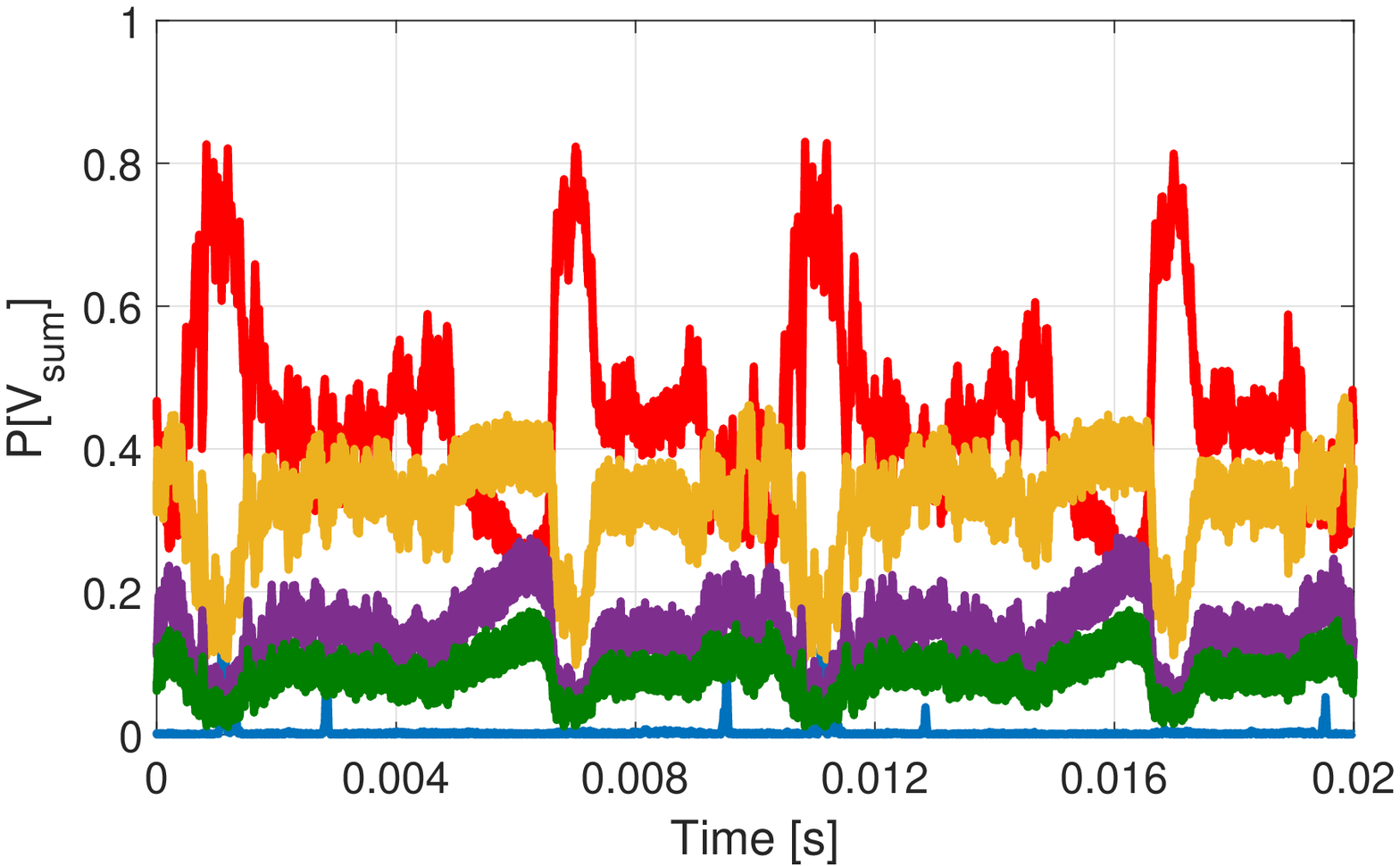} }}%
	\caption{Probability of a certain noise sample to reside in a specific voltage range. (a) $V_{sub} = V_{CH1}-V_{CH2}$ and (b) $V_{sum} = V_{CH1}+V_{CH2}$.}%
	\label{fig:noise_ex}%
\end{figure}
A full band analysis provides parameters such as voltage picks, averages and global statistics. While, the sub-bands, namely the CENELEC and above CENELEC bands, analysis gives insights on the signal correlation between the MIMO channels. In the following, some results on the measurement campaign are reported.
With the full bandwidth signal, the sum ($V_{sum} = V_{CH1}+V_{CH2}$) and difference ($V_{sub} = V_{CH1}-V_{CH2}$) of the two MIMO channels is performed in order to emphasize certain noise characteristics. Fig. \ref{fig:noise_ex}, for example shows the probability of a certain noise sample in the mains cycle of 20 ms to reside in a specific voltage range. The probability is computed as
\begin{align}
P[V_{sum}] &= P[V_1<V_{sum}<V_2 | n = n_0], \\
V_{sum} &= V_{sum}(nT_s + kT) \notag
\label{eq:prob}%
\end{align}
where, $n$ is the sample index, $T_s$ is the sampling period, $k$ is the mains period index, $T$ is the mains period, i.e. 20 ms, while, $V_1$ and $V_2$ are the voltage range limits.
These graphs highlight that the probability to find a certain noise level is not the same for every time slot. Specifically, (b) shows that voltage spikes over 316 mV have around 10\% of probability only to occur in brief time window (around 150 $\mu s$) and are located in specific time slots. Moreover, these graphs show that on average the signal $V_{sum}$ exhibits larger peaks with higher probability w.r.t. $V_{sub}$.

\begin{figure}
	\centering
	\subfloat[]{{\includegraphics[width=8.5cm,height=5cm]{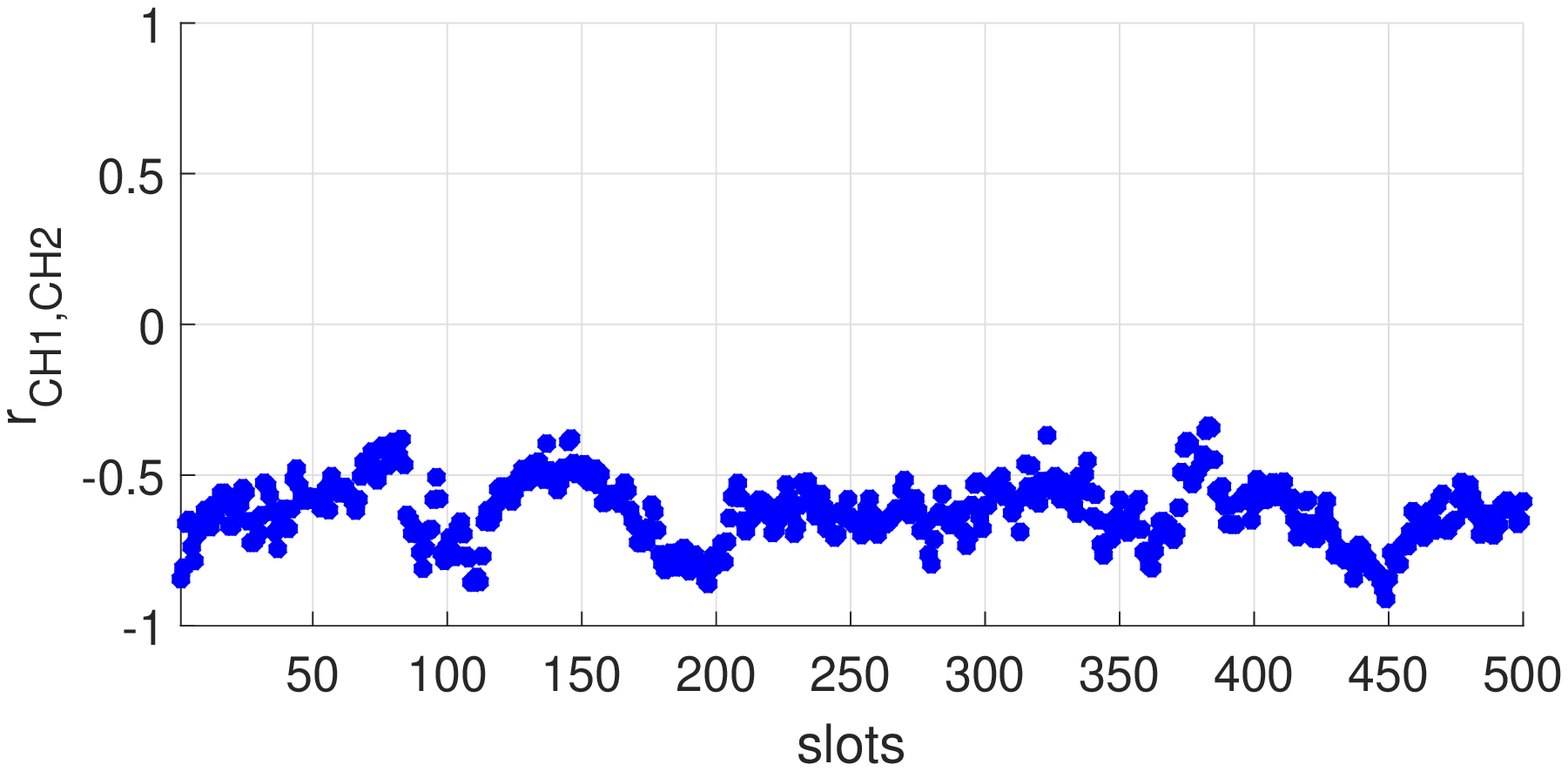} }}%
	\subfloat[]{{\includegraphics[width=8.5cm,height=5cm]{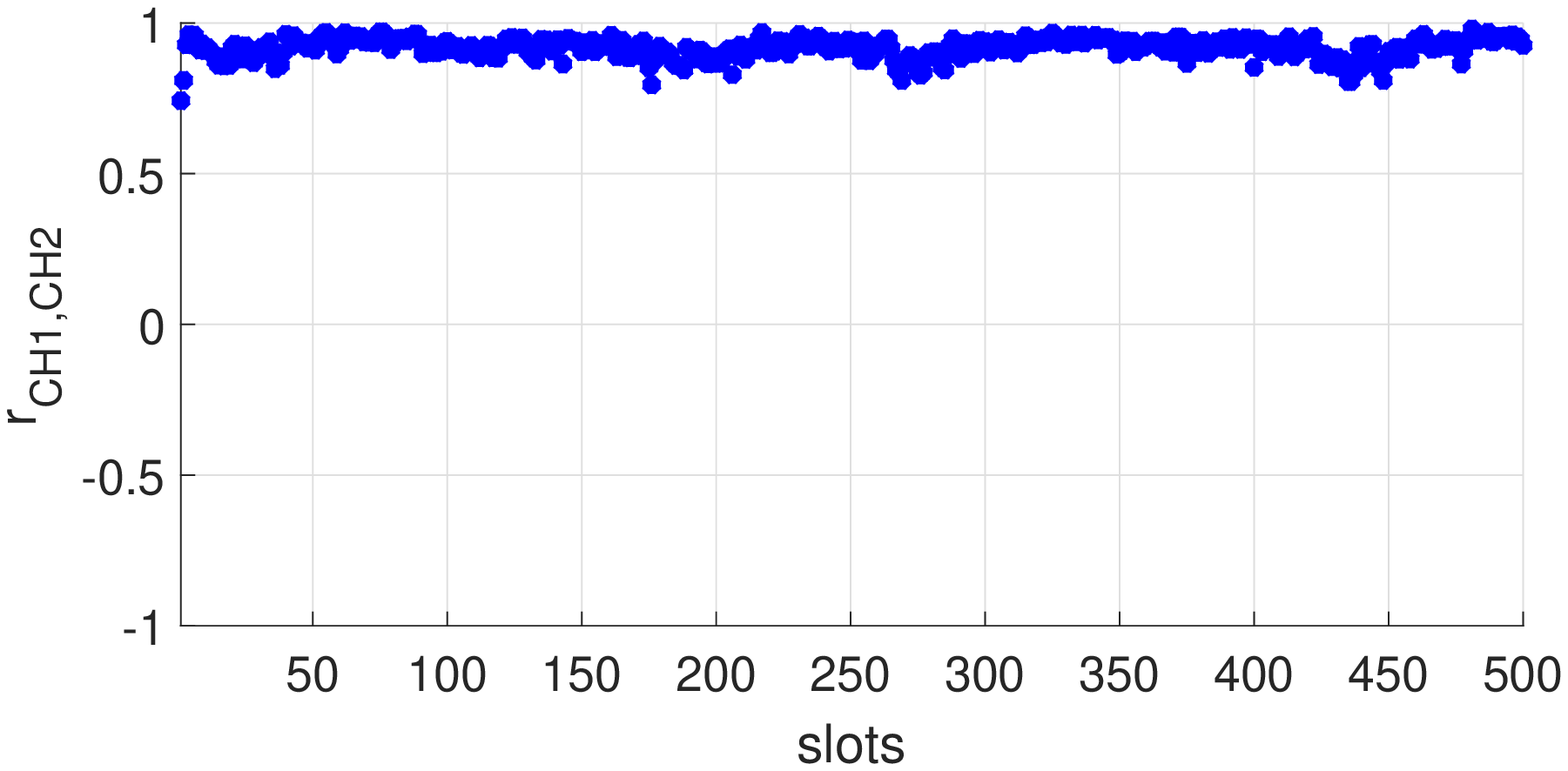} }}%
	\caption{Average cross-correlation coefficient with lag = 0	between CH1 and CH2 of the filtered noise signal in the CENELEC	D band. (a) and (b) are computed from noise acquired in two different outlets.}%
	\label{fig:xcorr_ex}%
\end{figure}

On the other hand, the sub-band analysis allows to better understand the relation between MC noise. In order to quantify this relation, the average cross-correlation of the channels is performed as explained in \cite{11}. Fig. \ref{fig:xcorr_ex} is an example of this computation. The graphs show in general a high correlation between the two channels. Fig. \ref{fig:xcorr_ex} (b) shows very high cross-correlation for more than 80\% of the time slots. While, Fig.  \ref{fig:xcorr_ex} (a) shows very high cross-correlation only in certain time windows. 

\section{Conclusion}
The results have shown that the noise not only has impulsive time variant nature, but it presents high correlation and in many cases determinism among multiple conductors. This is due to the symmetric multi-conductor geometries and the coupling of noise in such conductors. Moreover, since the time-location of noise spikes has high probability only in specific time intervals, physical layer algorithms can easily exploit this characteristic to mitigate noise and improve the reliability of the communication. The overall outcomes of this work offer new insights on the MC noise and useful information to build reliable physical layer protocols and noise canceling algorithms.

\end{document}